\documentclass[twocolumn,showpacs,amsmath,amssymb]{revtex4}
\usepackage{graphicx}
\usepackage{amsmath}
\usepackage{longtable}
\usepackage{afterpage}
\usepackage{rotating}

\begin{document}

\title{Measurement of magic wavelengths for the $^{40}$Ca$^+$ clock transition}

\author{Pei-Liang Liu$^{1,2,3}$, Yao Huang$^{1, 2}$, Wu Bian$^{1, 2, 3}$, Hu Shao$^{1, 2, 3}$, Hua Guan$^{1, 2*}$, \footnotetext{* Email: guanhua@wipm.ac.cn} Yong-Bo
Tang$^{1, 4}$, Cheng-Bin Li$^{1, 2}$, J. Mitroy$^5$, and Ke-Lin
Gao$^{1, 2\dag}$ \footnotetext{\dag \,\,Email: klgao@wipm.ac.cn} }
\affiliation {$^1$State Key Laboratory of Magnetic Resonance and
Atomic and Molecular Physics, Wuhan Institute of Physics and
Mathematics, Chinese Academy of Sciences, Wuhan 430071, China}
\affiliation {$^2$Key Laboratory of Atomic Frequency Standards,
Wuhan Institute of Physics and Mathematics, Chinese Academy of
Sciences, Wuhan 430071, China} \affiliation {$^3$University of
Chinese Academy of Sciences, Beijing 100049, China} \affiliation
{$^4$Department of Physics, Wuhan University, Wuhan 430072, China}
\affiliation {$^5$School of Engineering, Charles Darwin University, Darwin, NT 0909, Australia}

\date{\today}

\begin{abstract}
We demonstrate experimentally the existence of magic wavelengths near 396 nm and determine the ratio of the oscillator strengths for a single trapped ion. For the first time, two magic wavelengths for the $^{40}$Ca$^+$ clock transition are measured simultaneously with high precision. By tuning a laser to an intermediate wavelength between two transitions  $4s_{1/2} \to 4p_{1/2}$ and $4s_{1/2} \to 4p_{3/2}$ in $^{40}$Ca$^+$, the sensitivity of the clock transition Stark shift to the oscillator strengths for the resonance transition is greatly enhanced. With the measured magic wavelengths, we further determine the ratio of the oscillator strengths with the deviation less than 0.5\%. Our experimental method may be applied to measure magic wavelengths for other ion clock transitions, which may pave the way for building all-optical trapped ion clocks.
\end{abstract}

\pacs{31.15.ac, 31.15.ap, 34.20.Cf} \maketitle

 The magic wavelength for an atomic transition is a wavelength for which the differential ac Stark shift vanishes~\cite{takamoto05a,ye08a,barber08a,yi11a,mitroy10a}. The existence of magic wavelength enables independent control of internal hyperfine-spin and external center-of-mass motions of atoms (including neutral atoms and atomic ions). Precision measurements of magic wavelengths in atoms are very important in studies of atomic structure. For example, a measurement of the line strength ratio can bring a new perspective for determining accurate transition matrix elements, that are important in testing the atomic structure theories, as well as the reliability of a model used in interpreting atomic parity nonconservation studies~\cite{Derevianko00a,Sahoo06a,Porsev09a}. The oscillator strength, which is directly related to the line strength, can be derived from the magic wavelength measurements. Furthermore, it is an important topic in atomic physics, which can be critical in astrophysical data analysis~\cite{kurucz11a}. Knowledge of important oscillator strengths, and polarizabilities for the two states associated with a clock transition in an atom or ion is essential to correct the black-body radiation shift. A similar concept is the tune-out wavelength~\cite{leblanc07a}, at which the dynamic polarizability of the concerned atomic state is zero. Recently, by measuring the tune-out wavelengths, the line strength ratios have been derived for neutral potassium and rubidium~\cite{herold12a, holmgren12a}.

In addition, magic wavelengths provide extensive applications in quantum state engineering and precision frequency metrology~\cite{ye08a}. Magic wavelengths in neutral atomic systems have been measured in several experiments~\cite{yi11a,brusch06a,takamoto09a,lemke09a,ludlow08a}. The optical dipole trap at the magic wavelength can eliminate the first-order Stark shift so that the systematic uncertainties can be reduced. Atomic clocks based on neutral atoms trapped in the magic wavelength optical lattices are new trend of development for optical clocks~\cite{takamoto05a,bloom14a,margolis09a,hinkley13a}. Recently, all-optical trapping of ions has been demonstrated~\cite{Krasnov14a,Enderlein12a} and it is thus meaningful to explore the possibility of trapping ions using magic wavelength optical lattices. With the all-optical trapping technique, an ion clock can be built with better performance. Therefore, the demonstration of magic wavelengths for ion clock transitions is a milestone for establishing all-optical trapped ion clocks.

In this Letter, for the first time, two magic wavelengths for the $^{40}$Ca$^+$ clock-transition are reported and a novel application of magic wavelengths in determining oscillator strength ratio is shown. As the $^{40}$Ca$^+$ optical clock is a typical ion optical clock, our method for measuring magic wavelengths can be applied to other similar systems, such as Sr$^+$, In$^+$, Hg$^+$, Al$^+$, and Ba$^+$, which have been chosen as candidates for building optical clocks~\cite{Dube14a,Barwood14a,Hayasaka12a,wang07a,Rosenband08a,Kleczewski12a}.

\begin{figure}[th]
\includegraphics[width=6.0cm,angle=0]{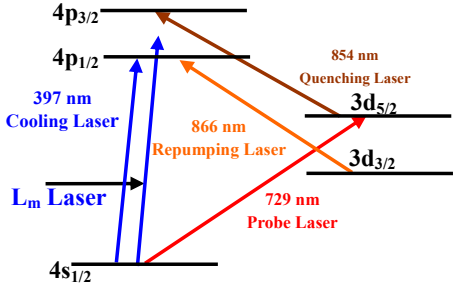}
\caption{(color online) partial energy level diagram of $^{40}$Ca$^+$.
 } \label{fig1}
\vspace{0.1cm}
\end{figure}

In Fig.~\ref{fig1}, we show the involved energy levels of $^{40}$Ca$^+$. The experiment introduced here uses a laser to apply an ac Stark shift to the $^{40}$Ca$^+$ $4s_{1/2} \to 3d_{5/2}$ ion clock transition for the $|m_j| = 1/2$ and $|m_j| = 3/2$ magnetic sub-level of the 3$d_{5/2}$ state. Measurements of the ac Stark shifts at different frequencies allow the determination of the magic wavelengths near 396 nm. There are two magic wavelengths $\lambda_{m_j=+1/2}$ and $\lambda_{m_j=-1/2}$ corresponding to the magnetic sub-states $m_j = \pm 1/2$ of $3d_{5/2}$ level. The magic wavelength $\lambda_{|m_j|=1/2}$ is the average of $\lambda_{m_j=+1/2}$ and $\lambda_{m_j=-1/2}$, similarly to $\lambda_{|m_j|=3/2}$. The measured magic wavelengths near 396 nm (L$_m$) lie between the resonant $4s_{1/2} \to 4p_{1/2}$ and $4s_{1/2} \to 4p_{3/2}$ transitions, as shown in Fig.~\ref{fig1}. Theoretical calculations indicated that these magic wavelengths were very sensitive to the ratio of the $4s_{1/2} \to 4p_{3/2}$ to $4s_{1/2} \to 4p_{1/2}$ oscillator strengths~\cite{tang13a}. The $^{40}$Ca$^+$ resonant oscillator strength cannot be determined with a single measurement since the
$4p_J$ states can decay to either the $4s_{1/2}$ ground state or the
$3d_J$ excited states. Besides the $4p_J$ lifetimes, one also
needs the branching ratio for transitions to the $4s_{1/2}$ and
$3d_J$ states~\cite{gallagher67a,gosselin88a,jin93a,gerritsma08a} since the $4p_J \to 3d_J$ transitions make
a contribution of about 6\% to the lifetimes~\cite{tang13a,gerritsma08a,safronova11a}. One of the advantages of the magic wavelength approach is that
the contribution to the polarizability from the $4s_{1/2} \to 4p_J$
transitions at the magic wavelength is three orders of
magnitude larger than the contribution from other transitions.

For an ion in a single mode laser field, the energy shift of a given
atomic state $a$ can be written as~\cite{takamoto09a}
\begin{eqnarray}
\Delta E_a = -\alpha_a(\omega) {\it{I}} - \beta_a(\omega) {\it{I}}^2 + O({\it{I}}^3),
\end{eqnarray}
where $\alpha_a(\omega)$ and $\beta_a(\omega)$ are the dynamic dipole polarizability and hyperpolarizability, respectively. Here, ${\it{I}}$ is the power density of the laser, and
$O({\it{I}}^3)$ represents the remaining high-order Stark shift.For the $^{40}$Ca$^+$ optical clock, one of major experimental
concerns is the frequency shift of the clock transition caused by electromagnetic radiation, which can be written as
\begin{equation}
h\Delta \nu  = \Delta E_d(\omega)-\Delta E_s(\omega)=-\Delta
\alpha(\omega){\it{I}}-\Delta
\beta(\omega){\it{I}}^2+\Delta O({\it{I}}^3),
\end{equation}
where $\Delta \alpha(\omega)$ and $\Delta \beta(\omega)$ are the differential
dipole-polarizability and hyperpolarizability, respectively. At the
magic wavelength $\lambda_{m_j}$ ($\lambda_{m_j}=c/\omega_{m_j}$ where c is the speed of light in vacuum), $\Delta \alpha(\omega)=0$. $\Delta \beta(\omega)$ is generally nonzero. Under the weak intensity limit, the contributions from the hyperpolarizability and the remaining higher-order terms are several orders of magnitude smaller than $\Delta \alpha(\omega){\it{I}}$ and can be ignored. Using ion optical
clock techniques, the differential light shift $\Delta\nu$ can be
measured accurately and therefore the magic wavelength $\lambda_{m_j}$ can be determined with
high precision.

\begin{figure}[th]
\includegraphics[width=8.4cm,angle=0]{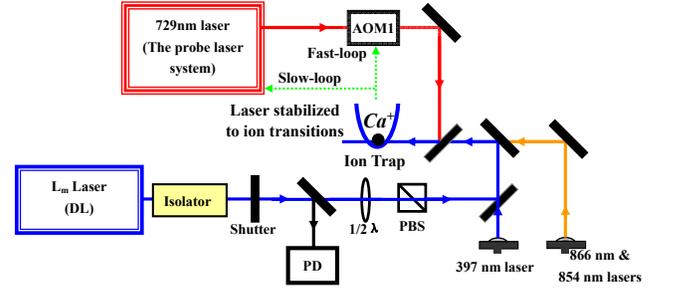}
\caption{(color online) Overview of magic wavelength measurement
setup. DL: diode
laser; AOM: acousto-optic modulator; 1/2$\lambda$: half wave plate;
PD: photo diode; PBS: polarized beam splitter.
 } \label{fig2}
\vspace{0.1cm}
\end{figure}

A sketch of the experimental setup for the measurement of magic wavelengths is shown in Fig.~\ref{fig2}. The whole system is composed of two main parts, one is the optical clock based on single trapped  $^{40}$Ca$^+$, and another one is the L$_m$ system for measuring the light shift of the clock transition.

The detail of the optical clock has been described in previous work~\cite{huang11a,huang12a}. A 729 nm probe laser was locked to an ultra-stable, high finesse cavity mounted on a vibration isolation platform (TS-140) by the Pound-Drever-Hall method, and an acousto-optic modulator is used to cancel the slow linear drift of the reference cavity.

The L$_m$  laser used in the experiment is frequency stabilized using a transfer cavity referenced to the 729 nm probe laser, and the long-term drift is reduced to less than 10 MHz within 4 hours. An unpolarized beam splitter (BS) is used to split a part of light for monitoring the laser power, which is 700 $\mu$W with a jitter of 3 $\mu$W. The power meter used in the experiment is a commercial power meter (S120VC, Thorlabs Inc.). The powers of the incident and output beams of the L$_m$ laser are monitored simultaneously. And they agree with each other very well on the jitter of laser power. The power of the L$_m$ laser into the trap is 731(4) $\mu$W and the waist radius of the beam is 203(5) $\mu$m during the measurement. A polarized beam splitter is placed in the light path before the ion-light interaction maintains the linear polarization of the L$_m$ laser. In this way, the linear polarization purity can reach 99.9\%, which can be derived by analyzing the polarization of the incident light and the transmission light of the L$_m$ laser.

In our work, a single $^{40}$Ca$^+$ ion is trapped in a miniature Paul trap and laser cooled to a few mK. The single ion's excess micromotion is minimized by adjusting the voltages of two compensation electrodes and two end-cap electrodes with the RF-photon correlation technique~\cite{berkeland98a} before performing any measurements. The $s-d$ clock transition splits symmetrically into ten Zeeman components around the zero-field line center~\cite{huang12a}. Then the probe laser is further referenced to the $^{40}$Ca$^+$ ion clock transitions by feeding back to the frequency of the acousto-optic modulator (AOM1) to compensate for changes of the magnetic field and probe the individual Zeeman transitions. In the experiment, the pulse sequences of 397 nm, 866 nm, 854 nm, and 729 nm lasers are similar to that used in the $^{40}$Ca$^+$ ion optical frequency standard~\cite{huang12a}. The pulse sequence of the L$_m$ laser is introduced to measure the light shift. The L$_m$ laser is off during the Doppler cooling period, and is on and off alternately during the probing stage to measure the light shift. The frequency values of AOM1 are recorded automatically every cycle by PC and the light shift caused by the L$_m$ laser beam can be measured by calculating the difference of two cycles with the L$_m$ laser on and off.

\begin{figure}[th]
\includegraphics[width=7.8cm,angle=0]{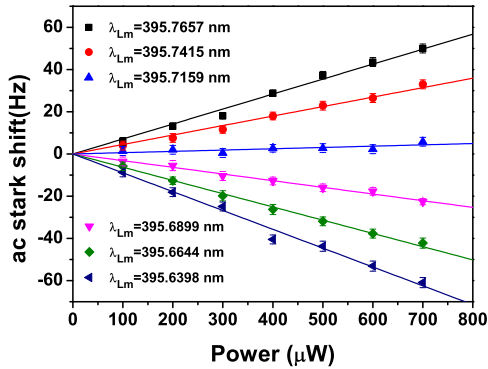}
\caption{(color online) The linear relation between the incident power of the L$_m$ laser and the ac Stark shift at different power strengths. All measurements correspond to the $^{40}$Ca$^+$ $4s_{1/2} \to 3d_{5/2}$ ion clock transition for the $m_j = +1/2$ magnetic sublevel of the $3d_{5/2}$ state .
 } \label{fig3}
\vspace{0.1cm}
\end{figure}

\begin{figure}[th]
\includegraphics[width=7.8cm,angle=0]{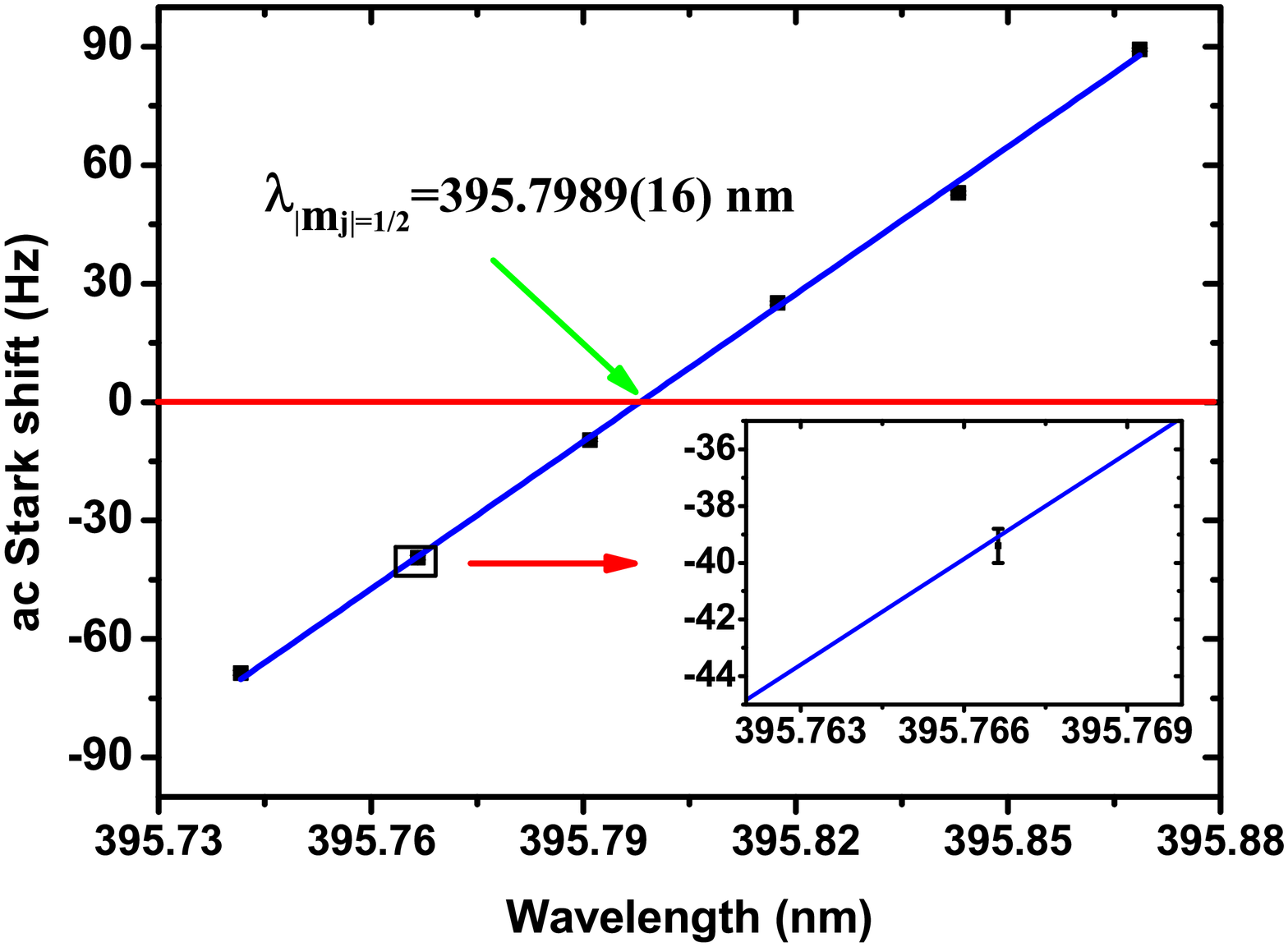}
\caption{(color online) The ac Stark shifts at different laser
wavelengths. The power of the Lm laser is 700 $\mu$W. Each data point represents 2000 s of experimental data. The
blue solid line is the linear fit to the data. The zero ac Stark
shift wavelength is identified as $\lambda_{m_j}$.The inserted figure shows the magnification for one measurement point.
 } \label{fig4}
\vspace{0.1cm}
\end{figure}

The ac Stark shifts within 0.2 nm around the magic wavelength $\lambda_{m_j}$ was investigated. Six fixed wavelengths of the L$_m$ laser were chosen and the ac Stark shifts were measured at each wavelength by switching the L$_m$ laser on/off . In order to study the relation between the incident power of the L$_m$ laser and the ac Stark shift, we measured the ac Stark shift of the $^{40}$Ca$^+$ $4s_{1/2} \to 3d_{5/2}$ ion clock transition for the $m_j = +1/2$ magnetic sublevel of the $3d_{5/2}$ state at different powers. The results are shown in Fig. ~\ref{fig3}. One can find a linear dependence on the power, which indicates that the quadratic and higher-order terms of Eq. (2) can be neglected in our experiment. The incident power of the L$_m$ laser on the ion probably changes when tuning the wavelength, since the laser beam direction may change slightly. Therefore the power was calibrated to ensure that it is identical within 4\% at all six wavelengths. The measured six points are fitted linearly and the magic wavelength is obtained. Fig.~\ref{fig4} shows one measurement of $\lambda_{|m_j|=1/2}$.

\begin{figure}[th]
\includegraphics[width=7.2cm,angle=0]{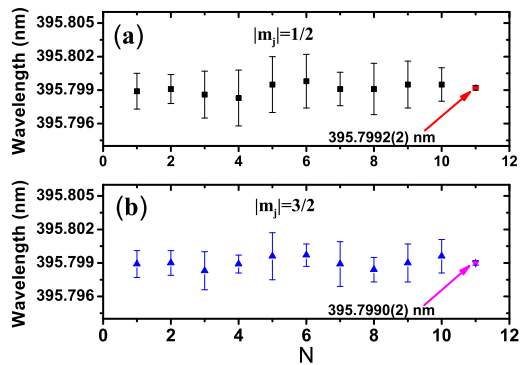}
\caption{(color online)(a) The 10 $\lambda_{|m_j|=1/2}$
measurements (black squares). (b) The 10 $\lambda_{|m_j|=3/2}$
measurements (blue triangle) under the same experimental conditions as the $\lambda_{|m_j|=1/2}$. The solid red circle
and the solid purple triangle indicate the weighted means and the
errors are the statistical errors of 10 $\lambda_{|m_j|=1/2}$ and $\lambda_{|m_j|=3/2}$ measurements.
 } \label{fig5}
\vspace{0.1cm}
\end{figure}

Due to the influence of the uncertainties from wave-meter measurement, the broadband spectral component and the power jitter of the L$_m$ laser, it is difficult to obtain the frequency difference by separate measurements of $|m_j| = 1/2$ and $|m_j| = 3/2$. Here a new measurement protocol was adopted. In the experiment, for each of the six wavelengths of the L$_m$ laser, we measured the ac Stark shifts for $|m_j| = 1/2$ and $|m_j| = 3/2$ of the 3$d_{5/2}$ state respectively. This procedure was repeated until all the six L$_m$ laser wavelengths had been considered. The L$_m$ laser's wavelength was tuned from 395.7 nm to 395.9 nm, then from 395.9 nm to 395.7 nm. Fig.~\ref{fig5} shows ten measurements for $\lambda_{|m_j|=1/2}$ and $\lambda_{|m_j|=3/2}$ respectively and the corresponding weighted means, resulting in $\lambda_{|m_j|=1/2} = 395.7992(2)$ nm and $\lambda_{|m_j|=3/2} = 395.7990(2)$ nm. The difference in value between $\lambda_{|m_j|=1/2}$ and $\lambda_{|m_j|=3/2}$ is 0.0002(6) nm, which agrees with the theoretical calculation~\cite{tang13a}.

To finalize the magic wavelength measurement of the L$_m$ laser,
systematic shifts must be considered and the corresponding corrections must be
applied to the above averaged frequencies. These shifts are due to the
broad spectral component, the light polarization, the second order
Doppler shift, the calibration of the wavemeter, $etc$. The error
budget is given in Table~\ref{tab1}.

\squeezetable
\begin{table}
\caption{\label{tab1} The uncertainty budget of the magic wavelength measurement. }
\begin{ruledtabular}
\begin {tabular}{lcc}
 Source of uncertainty  & Shift (pm) & Uncertainty (pm)\\
Broadband light         & 0          &  0.60  \\
Light polarization      & 0          & 0.01  \\
Second order Doppler shift and Stark shift  & 0.01   & 0.01  \\
Laser wavelength        & 0          & 0.06  \\
Statistical uncertainty & -          & 0.20  \\
\hline
  Total                 & 0.01       & 0.7  \\
\end{tabular}
\end{ruledtabular}
\end{table}
%

One of the major errors of the magic wavelength measurement
comes from the broad spectral component of L$_m$. To evaluate the
broad spectral component, a grating spectrometer (IHR550, HORIBA) was
used to analyze the laser spectrum. We found that more than 99\% of the
laser power was within the wavelength range of 0.03 nm and only
less than 1\% laser power was out of that range, corresponding to a
less than 0.0005 nm contribution to the uncertainty in $\lambda_{|m_j|=1/2}$
and $\lambda_{|m_j|=3/2}$~\cite{holmgren12a}. Next, the spectral
component in the range of 0.03 nm around the carrier was analyzed by observing
the beatnote of the L$_m$ lasers with another similar laser using a
spectrum analyzer. The spectrum mainly contained three components from which the ac Stark shift could be
estimated based on their relative weights and only an uncertainty of less than 0.0001 nm was obtained. Lights with
different polarizations can result in different ac Stark shifts.
Elliptical polarization will have scalar, vector and tensor light
shifts for atoms with $|m_F|>0$~\cite{deutsch10a}, which will change the
value of the magic wavelength. In our experiment, a polarized beam
splitter (PBS) was used to create a pure linear polarization, and the
ellipticity component was reduced to less than 0.1\% by analyzing the light
beam before and after the vacuum chamber. To evaluate the
contribution due to the non-linearly polarized component, the uncertainty with circularly polarized light
was measured and the wavelength difference with linearly and
circularly polarized light was less than 0.01 nm, thus there would only be
less than 0.0001 nm of uncertainty with less than 0.1\% ellipticity.

The L$_m$ laser could heat the ion or affect the efficiency of the
laser cooling, introducing a second order Doppler shift and a Stark
shift due to the increase of the energy of the ion thermal motion or micromotion.
To estimate the second-order Doppler shift, the ion temperature was
measured by monitoring the intensity of secular sidebands, together with
the measurements of the micromotion using rf-photon correlation
method~\cite{berkeland98a} with and without the L$_m$ laser, resulting in an uncertainty of 0.00001 nm. The L$_m$ laser wavelength after frequency stabilization was monitored by a wavemeter (High Finesse WS-7) with an absolute accuracy of 60 MHz after the
calibration using the clock laser. Thus the uncertainty from the calibration of the wavemeter is within 0.00006 nm.
Our final determined magic wavelengths are 395.7992(7) nm and 395.7990(7) nm, located in the spin-orbit energy gap of the 4p state.

The dynamic Stark shift is strongly dominated by the large and opposite
polarizability contributions from the 4$p_{1/2}$ and 4$p_{3/2}$
states~\cite{tang13a,herold12a,arora07a}. The contributions of the
3$d_{5/2}$ polarizabilities are typically small in magnitude at this
wavelength. The dynamic Stark shift can be written as
\begin{eqnarray}
0&=&\alpha_{4s_{1/2}}(\omega_{m_j}) -
\alpha_{3d_{5/2}}(\omega_{m_j})\nonumber
\\
&\cong&
\frac{f(4s_{1/2}\to 4p_{1/2})}{\epsilon^2_{4s_{1/2}\to 4p_{1/2}}-\omega_{m_j}^2}
 +
\frac{f(4s_{1/2}\to 4p_{3/2})}{\epsilon^2_{4s_{1/2} \to 4p_{3/2}}-\omega_{m_j}^2}
+ \Delta, \nonumber \\ \label{eq}
\end{eqnarray}
where $\epsilon$ is the transition energy, $f$ is the oscillator strength, and
$\Delta$ consists of the remaining terms in the 4$s_{1/2}$
polarizability, as well as the smaller 3$d_{5/2}$ polarizabilities that are estimated to be 2.95 a$^3_0$ for the 3$d_{m_j=1/2}$ state and 0.31 a$^3_0$
for the 3$d_{m_j=3/2}$ state. The value of $f(4s_{1/2} \to 4p_{1/2})$ was determined theoretically by us to be 0.3171.

Using the energies of the $4s_{1/2}$, $4p_{1/2}$, and $4p_{3/2}$ states from NIST Database~\cite{nistasd500} and the present measured magic wavelength $\omega_{m_j}$, we can obtain
the oscillator strength ratio $R_f$ according to Eq.~(\ref{eq})
\begin{eqnarray}
R_f=\frac{f(4s_{1/2}-4p_{3/2})}{f(4s_{1/2}-4p_{1/2})}=2.027(5),
\end{eqnarray}
and the line strength ratio $R_s$ is
\begin{eqnarray}
R_s=\frac{ \big|\langle 4s \| D \| 4p_{3/2}\rangle \big|^2
}{\big|\langle 4s \| D \| 4p_{1/2}\rangle \big|^2}=2.009(5).
\end{eqnarray}
A change in $\Delta$ by 2.0 a$^3_0$ will result in a change in the derived $R_s$
by 0.001. Changes in the oscillator strengths of the
background transitions of more than 5\% would be needed to change
$\Delta$ by 2.0 a$^3_0$ and the uncertainty estimates in $R_s$ and
$R_f$ allows for this. The previously estimated line strength ratios of 2.001~\cite{safronova11a} and 2.0014~\cite{tang13a} were based onthe relativistic all-order many-body
perturbation theory and the relativistic semi-empricial potential, respectively.
Since the results for $\omega_{m_j=1/2}$ and $\omega_{m_j=3/2}$ are very close to each other, the final ratios of $R_f = 2.027(5)$ and $R_s = 2.009(5)$ include both the experimental and theoretical uncertainties.

In summary, two magic wavelengths of $\lambda_{|m_j|=1/2} =
395.7992(7)$ nm and $\lambda_{|m_j|=3/2} = 395.7990(7)$ nm of the $^{40}$Ca$^+$ clock transition have been measured with the accuracy of 2 ppm.
Our experiment is the first demonstration of finding magic wavelengths in an ion optical clock system. The oscillator-strength ratio and the
line-strength ratio for the transitions $f(4s_{1/2} \to 4p_{3/2, 1/2})$ have been determined to be 2.027(5) and 2.009(5). At present,
the broadband spectrum of the L$_m$ laser and statistical error were the largest contributors to the total systematic uncertainty. These errors can be reduced by introducing a cavity for mode selection to the L$_m$ laser and by improving the power stabilization, respectively. An order of magnitude improvement over the currently determined magic wavelengths is therefore achievable in future work.

\begin{acknowledgments}

We thank Ting-Yun Shi and Li-Yan Tang for valuable suggestion. Thank Z. -
C. Yan (UNB), J. Ye (JILA), H. Klein (NPL), C. Lee (SYSU), X. Guan, and J. Chen (IAPCM) for fruitful discussions. And thank X. Huang, H. Shu, H. Fan, B. Guo, Q. Liu, W. Qu, J. Cao, and B. Ou for the early experiments. This work is supported by the National Basic Research Program of China (2012CB821301), the National Natural Science Foundation of China (11474318, 91336211 and 11034009) and Chinese Academy of Sciences.

\end{acknowledgments}



\begin{thebibliography}{38}
\expandafter\ifx\csname natexlab\endcsname\relax\def\natexlab#1{#1}\fi
\expandafter\ifx\csname bibnamefont\endcsname\relax
  \def\bibnamefont#1{#1}\fi
\expandafter\ifx\csname bibfnamefont\endcsname\relax
  \def\bibfnamefont#1{#1}\fi
\expandafter\ifx\csname citenamefont\endcsname\relax
  \def\citenamefont#1{#1}\fi
\expandafter\ifx\csname url\endcsname\relax
  \def\url#1{\texttt{#1}}\fi
\expandafter\ifx\csname urlprefix\endcsname\relax\def\urlprefix{URL }\fi
\providecommand{\bibinfo}[2]{#2}
\providecommand{\eprint}[2][]{\url{#2}}

\bibitem[{\citenamefont{{Takamoto} et~al.}(2005)\citenamefont{{Takamoto},
  {Hong}, {Higashi}, and {Katori}}}]{takamoto05a}
\bibinfo{author}{\bibfnamefont{M.}~\bibnamefont{{Takamoto}}},
  \bibinfo{author}{\bibfnamefont{F.-L.} \bibnamefont{{Hong}}},
  \bibinfo{author}{\bibfnamefont{R.}~\bibnamefont{{Higashi}}},
  \bibnamefont{and} \bibinfo{author}{\bibfnamefont{H.}~\bibnamefont{{Katori}}},
  \bibinfo{journal}{Nature} \textbf{\bibinfo{volume}{435}},
  \bibinfo{pages}{321} (\bibinfo{year}{2005}).

\bibitem[{\citenamefont{{Ye} et~al.}(2008)\citenamefont{{Ye}, {Kimble}, and
  {Katori}}}]{ye08a}
\bibinfo{author}{\bibfnamefont{J.}~\bibnamefont{{Ye}}},
  \bibinfo{author}{\bibfnamefont{H.~J.} \bibnamefont{{Kimble}}},
  \bibnamefont{and} \bibinfo{author}{\bibfnamefont{H.}~\bibnamefont{{Katori}}},
  \bibinfo{journal}{Science} \textbf{\bibinfo{volume}{320}},
  \bibinfo{pages}{1734} (\bibinfo{year}{2008}).

\bibitem[{\citenamefont{{Barber} et~al.}(2008)\citenamefont{{Barber},
  {Stalnaker}, {Lemke}, {Poli}, {Oates}, {Fortier}, {Diddams}, {Hollberg},
  {Hoyt}, {Taichenachev} et~al.}}]{barber08a}
\bibinfo{author}{\bibfnamefont{Z.~W.} \bibnamefont{{Barber}}},
  \bibinfo{author}{\bibfnamefont{J.~E.} \bibnamefont{{Stalnaker}}},
  \bibinfo{author}{\bibfnamefont{N.~D.} \bibnamefont{{Lemke}}},
  \bibinfo{author}{\bibfnamefont{N.}~\bibnamefont{{Poli}}},
  \bibinfo{author}{\bibfnamefont{C.~W.} \bibnamefont{{Oates}}},
  \bibinfo{author}{\bibfnamefont{T.~M.} \bibnamefont{{Fortier}}},
  \bibinfo{author}{\bibfnamefont{S.~A.} \bibnamefont{{Diddams}}},
  \bibinfo{author}{\bibfnamefont{L.}~\bibnamefont{{Hollberg}}},
  \bibinfo{author}{\bibfnamefont{C.~W.} \bibnamefont{{Hoyt}}},
  \bibinfo{author}{\bibfnamefont{A.~V.} \bibnamefont{{Taichenachev}}},
  \bibnamefont{et~al.}, \bibinfo{journal}{Phys.~Rev.~Lett.}
  \textbf{\bibinfo{volume}{100}}, \bibinfo{eid}{103002} (\bibinfo{year}{2008}).

\bibitem[{\citenamefont{{Yi} et~al.}(2011)\citenamefont{{Yi}, {Mejri},
  {McFerran}, {Le Coq}, and {Bize}}}]{yi11a}
\bibinfo{author}{\bibfnamefont{L.}~\bibnamefont{{Yi}}},
  \bibinfo{author}{\bibfnamefont{S.}~\bibnamefont{{Mejri}}},
  \bibinfo{author}{\bibfnamefont{J.~J.} \bibnamefont{{McFerran}}},
  \bibinfo{author}{\bibfnamefont{Y.}~\bibnamefont{{Le Coq}}}, \bibnamefont{and}
  \bibinfo{author}{\bibfnamefont{S.}~\bibnamefont{{Bize}}},
  \bibinfo{journal}{Phys.~Rev.~Lett.} \textbf{\bibinfo{volume}{106}},
  \bibinfo{eid}{073005} (\bibinfo{year}{2011}).

\bibitem[{\citenamefont{{Mitroy} et~al.}(2010)\citenamefont{{Mitroy},
  {Safronova}, and {Clark}}}]{mitroy10a}
\bibinfo{author}{\bibfnamefont{J.}~\bibnamefont{{Mitroy}}},
  \bibinfo{author}{\bibfnamefont{M.~S.} \bibnamefont{{Safronova}}},
  \bibnamefont{and} \bibinfo{author}{\bibfnamefont{C.~W.}
  \bibnamefont{{Clark}}}, \bibinfo{journal}{J.~Phys.~B}
  \textbf{\bibinfo{volume}{43}}, \bibinfo{eid}{202001} (\bibinfo{year}{2010}).

\bibitem[{\citenamefont{{Derevianko}}(2000)}]{Derevianko00a}
\bibinfo{author}{\bibfnamefont{A.}~\bibnamefont{{Derevianko}}},
  \bibinfo{journal}{\prl} \textbf{\bibinfo{volume}{85}}, \bibinfo{eid}{1618}
  (\bibinfo{year}{2000}).

\bibitem[{\citenamefont{{Sahoo} et~al.}(2006)\citenamefont{{Sahoo},
  {Chaudhuri}, {Das}, and {Mukherjee}}}]{Sahoo06a}
\bibinfo{author}{\bibfnamefont{B.~K.} \bibnamefont{{Sahoo}}},
  \bibinfo{author}{\bibfnamefont{R.~K.} \bibnamefont{{Chaudhuri}}},
  \bibinfo{author}{\bibfnamefont{B.~P.} \bibnamefont{{Das}}}, \bibnamefont{and}
  \bibinfo{author}{\bibfnamefont{D.}~\bibnamefont{{Mukherjee}}},
  \bibinfo{journal}{\prl} \textbf{\bibinfo{volume}{96}}, \bibinfo{eid}{163003}
  (\bibinfo{year}{2006}).

\bibitem[{\citenamefont{{Porsev} et~al.}(2009)\citenamefont{{Porsev}, {Beloy },
  and {Derevianko}}}]{Porsev09a}
\bibinfo{author}{\bibfnamefont{S.~G.} \bibnamefont{{Porsev}}},
  \bibinfo{author}{\bibfnamefont{K.}~\bibnamefont{{Beloy }}}, \bibnamefont{and}
  \bibinfo{author}{\bibfnamefont{A.}~\bibnamefont{{Derevianko}}},
  \bibinfo{journal}{\prl} \textbf{\bibinfo{volume}{102}}, \bibinfo{eid}{181601}
  (\bibinfo{year}{2009}).

\bibitem[{\citenamefont{{Kurucz}}(2011)}]{kurucz11a}
\bibinfo{author}{\bibfnamefont{R.~L.} \bibnamefont{{Kurucz}}},
  \bibinfo{journal}{Can.~J.~Phys.} \textbf{\bibinfo{volume}{89}},
  \bibinfo{pages}{417} (\bibinfo{year}{2011}).

\bibitem[{\citenamefont{{LeBlanc} and {Thywissen}}(2007)}]{leblanc07a}
\bibinfo{author}{\bibfnamefont{L.~J.} \bibnamefont{{LeBlanc}}}
  \bibnamefont{and} \bibinfo{author}{\bibfnamefont{J.~H.}
  \bibnamefont{{Thywissen}}}, \bibinfo{journal}{\pra}
  \textbf{\bibinfo{volume}{75}}, \bibinfo{eid}{053612} (\bibinfo{year}{2007}).

\bibitem[{\citenamefont{{Herold} et~al.}(2012)\citenamefont{{Herold}, {Vaidya},
  {Li}, {Rolston}, {Porto}, and {Safronova}}}]{herold12a}
\bibinfo{author}{\bibfnamefont{C.~D.} \bibnamefont{{Herold}}},
  \bibinfo{author}{\bibfnamefont{V.~D.} \bibnamefont{{Vaidya}}},
  \bibinfo{author}{\bibfnamefont{X.}~\bibnamefont{{Li}}},
  \bibinfo{author}{\bibfnamefont{S.~L.} \bibnamefont{{Rolston}}},
  \bibinfo{author}{\bibfnamefont{J.~V.} \bibnamefont{{Porto}}},
  \bibnamefont{and} \bibinfo{author}{\bibfnamefont{M.~S.}
  \bibnamefont{{Safronova}}}, \bibinfo{journal}{\prl}
  \textbf{\bibinfo{volume}{109}}, \bibinfo{eid}{243003} (\bibinfo{year}{2012}).

\bibitem[{\citenamefont{{Holmgren} et~al.}(2012)\citenamefont{{Holmgren},
  {Trubko}, {Hromada}, and {Cronin}}}]{holmgren12a}
\bibinfo{author}{\bibfnamefont{W.~F.} \bibnamefont{{Holmgren}}},
  \bibinfo{author}{\bibfnamefont{R.}~\bibnamefont{{Trubko}}},
  \bibinfo{author}{\bibfnamefont{I.}~\bibnamefont{{Hromada}}},
  \bibnamefont{and} \bibinfo{author}{\bibfnamefont{A.~D.}
  \bibnamefont{{Cronin}}}, \bibinfo{journal}{\prl}
  \textbf{\bibinfo{volume}{109}}, \bibinfo{eid}{243004} (\bibinfo{year}{2012}).

\bibitem[{\citenamefont{{Brusch} et~al.}(2006)\citenamefont{{Brusch}, {Le
  Targat}, {Baillard}, {Fouch{\'e}}, and {Lemonde}}}]{brusch06a}
\bibinfo{author}{\bibfnamefont{A.}~\bibnamefont{{Brusch}}},
  \bibinfo{author}{\bibfnamefont{R.}~\bibnamefont{{Le Targat}}},
  \bibinfo{author}{\bibfnamefont{X.}~\bibnamefont{{Baillard}}},
  \bibinfo{author}{\bibfnamefont{M.}~\bibnamefont{{Fouch{\'e}}}},
  \bibnamefont{and}
  \bibinfo{author}{\bibfnamefont{P.}~\bibnamefont{{Lemonde}}},
  \bibinfo{journal}{Phys.~Rev.~Lett.} \textbf{\bibinfo{volume}{96}},
  \bibinfo{eid}{103003} (\bibinfo{year}{2006}).

\bibitem[{\citenamefont{{Takamoto} et~al.}(2009)\citenamefont{{Takamoto},
  {Katori}, {Marmo}, {Ovsiannikov}, and {Pal'chikov}}}]{takamoto09a}
\bibinfo{author}{\bibfnamefont{M.}~\bibnamefont{{Takamoto}}},
  \bibinfo{author}{\bibfnamefont{H.}~\bibnamefont{{Katori}}},
  \bibinfo{author}{\bibfnamefont{S.~I.} \bibnamefont{{Marmo}}},
  \bibinfo{author}{\bibfnamefont{V.~D.} \bibnamefont{{Ovsiannikov}}},
  \bibnamefont{and} \bibinfo{author}{\bibfnamefont{V.~G.}
  \bibnamefont{{Pal'chikov}}}, \bibinfo{journal}{Phys.~Rev.~Lett.}
  \textbf{\bibinfo{volume}{102}}, \bibinfo{eid}{063002} (\bibinfo{year}{2009}).

\bibitem[{\citenamefont{{Lemke} et~al.}(2009)\citenamefont{{Lemke}, {Ludlow},
  {Barber}, {Fortier}, {Diddams}, {Jiang}, {Jefferts}, {Heavner}, {Parker}, and
  {Oates}}}]{lemke09a}
\bibinfo{author}{\bibfnamefont{N.~D.} \bibnamefont{{Lemke}}},
  \bibinfo{author}{\bibfnamefont{A.~D.} \bibnamefont{{Ludlow}}},
  \bibinfo{author}{\bibfnamefont{Z.~W.} \bibnamefont{{Barber}}},
  \bibinfo{author}{\bibfnamefont{T.~M.} \bibnamefont{{Fortier}}},
  \bibinfo{author}{\bibfnamefont{S.~A.} \bibnamefont{{Diddams}}},
  \bibinfo{author}{\bibfnamefont{Y.}~\bibnamefont{{Jiang}}},
  \bibinfo{author}{\bibfnamefont{S.~R.} \bibnamefont{{Jefferts}}},
  \bibinfo{author}{\bibfnamefont{T.~P.} \bibnamefont{{Heavner}}},
  \bibinfo{author}{\bibfnamefont{T.~E.} \bibnamefont{{Parker}}},
  \bibnamefont{and} \bibinfo{author}{\bibfnamefont{C.~W.}
  \bibnamefont{{Oates}}}, \bibinfo{journal}{Phys.~Rev.~Lett.}
  \textbf{\bibinfo{volume}{103}}, \bibinfo{eid}{063001} (\bibinfo{year}{2009}).

\bibitem[{\citenamefont{{Ludlow} et~al.}(2008)\citenamefont{{Ludlow},
  {Zelevinsky}, {Campbell}, {Blatt}, {Boyd}, {de Miranda}, {Martin}, {Thomsen},
  {Foreman}, {Ye} et~al.}}]{ludlow08a}
\bibinfo{author}{\bibfnamefont{A.~D.} \bibnamefont{{Ludlow}}},
  \bibinfo{author}{\bibfnamefont{T.}~\bibnamefont{{Zelevinsky}}},
  \bibinfo{author}{\bibfnamefont{G.~K.} \bibnamefont{{Campbell}}},
  \bibinfo{author}{\bibfnamefont{S.}~\bibnamefont{{Blatt}}},
  \bibinfo{author}{\bibfnamefont{M.~M.} \bibnamefont{{Boyd}}},
  \bibinfo{author}{\bibfnamefont{M.~H.~G.} \bibnamefont{{de Miranda}}},
  \bibinfo{author}{\bibfnamefont{M.~J.} \bibnamefont{{Martin}}},
  \bibinfo{author}{\bibfnamefont{J.~W.} \bibnamefont{{Thomsen}}},
  \bibinfo{author}{\bibfnamefont{S.~M.} \bibnamefont{{Foreman}}},
  \bibinfo{author}{\bibfnamefont{J.}~\bibnamefont{{Ye}}}, \bibnamefont{et~al.},
  \bibinfo{journal}{Science} \textbf{\bibinfo{volume}{319}},
  \bibinfo{pages}{1805} (\bibinfo{year}{2008}).

\bibitem[{\citenamefont{{Bloom} et~al.}(2014)\citenamefont{{Bloom},
  {Nicholson}, {Williams}, {Campbell}, {Bishof}, {Zhang}, {Zhang}, {Bromley},
  and {Ye}}}]{bloom14a}
\bibinfo{author}{\bibfnamefont{B.~J.} \bibnamefont{{Bloom}}},
  \bibinfo{author}{\bibfnamefont{T.~L.} \bibnamefont{{Nicholson}}},
  \bibinfo{author}{\bibfnamefont{J.~R.} \bibnamefont{{Williams}}},
  \bibinfo{author}{\bibfnamefont{S.~L.} \bibnamefont{{Campbell}}},
  \bibinfo{author}{\bibfnamefont{M.}~\bibnamefont{{Bishof}}},
  \bibinfo{author}{\bibfnamefont{X.}~\bibnamefont{{Zhang}}},
  \bibinfo{author}{\bibfnamefont{W.}~\bibnamefont{{Zhang}}},
  \bibinfo{author}{\bibfnamefont{S.~L.} \bibnamefont{{Bromley}}},
  \bibnamefont{and} \bibinfo{author}{\bibfnamefont{J.}~\bibnamefont{{Ye}}},
  \bibinfo{journal}{Nature} \textbf{\bibinfo{volume}{506}}, \bibinfo{pages}{71}
  (\bibinfo{year}{2014}).

\bibitem[{\citenamefont{{Margolis}}(2009)}]{margolis09a}
\bibinfo{author}{\bibfnamefont{H.~S.} \bibnamefont{{Margolis}}},
  \bibinfo{journal}{J.~Phys.~B} \textbf{\bibinfo{volume}{42}},
  \bibinfo{eid}{154017} (\bibinfo{year}{2009}).

\bibitem[{\citenamefont{{Hinkley} et~al.}(2013)\citenamefont{{Hinkley},
  {Sherman}, {Phillips}, {Schioppo}, {Lemke}, {Beloy}, {Pizzocaro}, {Oates},
  and {Ludlow}}}]{hinkley13a}
\bibinfo{author}{\bibfnamefont{N.}~\bibnamefont{{Hinkley}}},
  \bibinfo{author}{\bibfnamefont{J.~A.} \bibnamefont{{Sherman}}},
  \bibinfo{author}{\bibfnamefont{N.~B.} \bibnamefont{{Phillips}}},
  \bibinfo{author}{\bibfnamefont{M.}~\bibnamefont{{Schioppo}}},
  \bibinfo{author}{\bibfnamefont{N.~D.} \bibnamefont{{Lemke}}},
  \bibinfo{author}{\bibfnamefont{K.}~\bibnamefont{{Beloy}}},
  \bibinfo{author}{\bibfnamefont{M.}~\bibnamefont{{Pizzocaro}}},
  \bibinfo{author}{\bibfnamefont{C.~W.} \bibnamefont{{Oates}}},
  \bibnamefont{and} \bibinfo{author}{\bibfnamefont{A.~D.}
  \bibnamefont{{Ludlow}}}, \bibinfo{journal}{Science}
  \textbf{\bibinfo{volume}{341}}, \bibinfo{pages}{1215} (\bibinfo{year}{2013}).

\bibitem[{\citenamefont{{Krasnov} and {Kamenshchikov}}(2014)}]{Krasnov14a}
\bibinfo{author}{\bibfnamefont{I.~V.} \bibnamefont{{Krasnov}}}
  \bibnamefont{and} \bibinfo{author}{\bibfnamefont{L.~P.}
  \bibnamefont{{Kamenshchikov}}}, \bibinfo{journal}{Optics~Commun.}
  \textbf{\bibinfo{volume}{312}}, \bibinfo{eid}{192} (\bibinfo{year}{2014}).

\bibitem[{\citenamefont{{Enderlein} et~al.}(2012)\citenamefont{{Enderlein},
  {Huber}, {Schneider}, and {Schaetz}}}]{Enderlein12a}
\bibinfo{author}{\bibfnamefont{M.}~\bibnamefont{{Enderlein}}},
  \bibinfo{author}{\bibfnamefont{T.}~\bibnamefont{{Huber}}},
  \bibinfo{author}{\bibfnamefont{C.}~\bibnamefont{{Schneider}}},
  \bibnamefont{and}
  \bibinfo{author}{\bibfnamefont{T.}~\bibnamefont{{Schaetz}}},
  \bibinfo{journal}{\prl} \textbf{\bibinfo{volume}{109}}, \bibinfo{eid}{233004}
  (\bibinfo{year}{2012}).

\bibitem[{\citenamefont{{Dub{\'e}} et~al.}(2014)\citenamefont{{Dub{\'e}},
  {Madej}, {Tibbo}, and {Bernard}}}]{Dube14a}
\bibinfo{author}{\bibfnamefont{P.}~\bibnamefont{{Dub{\'e}}}},
  \bibinfo{author}{\bibfnamefont{A.~A.} \bibnamefont{{Madej}}},
  \bibinfo{author}{\bibfnamefont{M.}~\bibnamefont{{Tibbo}}}, \bibnamefont{and}
  \bibinfo{author}{\bibfnamefont{J.~E.} \bibnamefont{{Bernard}}},
  \bibinfo{journal}{\prl} \textbf{\bibinfo{volume}{112}}, \bibinfo{eid}{173002}
  (\bibinfo{year}{2014}).

\bibitem[{\citenamefont{{Barwood} et~al.}(2014)\citenamefont{{Barwood},
  {Huang}, {Klein}, {Johnson}, {King}, {Margolis}, {Szymaniec}, and
  {Gill}}}]{Barwood14a}
\bibinfo{author}{\bibfnamefont{G.~P.} \bibnamefont{{Barwood}}},
  \bibinfo{author}{\bibfnamefont{G.}~\bibnamefont{{Huang}}},
  \bibinfo{author}{\bibfnamefont{H.~A.} \bibnamefont{{Klein}}},
  \bibinfo{author}{\bibfnamefont{L.~A.~M.} \bibnamefont{{Johnson}}},
  \bibinfo{author}{\bibfnamefont{S.~A.} \bibnamefont{{King}}},
  \bibinfo{author}{\bibfnamefont{H.~S.} \bibnamefont{{Margolis}}},
  \bibinfo{author}{\bibfnamefont{K.}~\bibnamefont{{Szymaniec}}},
  \bibnamefont{and} \bibinfo{author}{\bibfnamefont{P.}~\bibnamefont{{Gill}}},
  \bibinfo{journal}{\pra} \textbf{\bibinfo{volume}{89}},
  \bibinfo{eid}{050501(R)} (\bibinfo{year}{2014}).

\bibitem[{\citenamefont{{Hayasaka}}(2012)}]{Hayasaka12a}
\bibinfo{author}{\bibfnamefont{K.}~\bibnamefont{{Hayasaka}}},
  \bibinfo{journal}{Appl.~Phys.~B} \textbf{\bibinfo{volume}{107}},
  \bibinfo{eid}{965} (\bibinfo{year}{2012}).

\bibitem[{\citenamefont{{Wang} et~al.}(2007)\citenamefont{{Wang}, {Dumke},
  {Liu}, {Stejskal}, {Zhao}, {Zhang}, {Lu}, {Wang}, {Becker}, and
  {Walther}}}]{wang07a}
\bibinfo{author}{\bibfnamefont{H.}~\bibnamefont{{Wang}}, \bibfnamefont{Y}},
  \bibinfo{author}{\bibfnamefont{R.}~\bibnamefont{{Dumke}}},
  \bibinfo{author}{\bibfnamefont{T.}~\bibnamefont{{Liu}}},
  \bibinfo{author}{\bibfnamefont{A.}~\bibnamefont{{Stejskal}}},
  \bibinfo{author}{\bibfnamefont{N.}~\bibnamefont{{Zhao}}, \bibfnamefont{Y}},
  \bibinfo{author}{\bibfnamefont{J.}~\bibnamefont{{Zhang}}},
  \bibinfo{author}{\bibfnamefont{H.}~\bibnamefont{{Lu}}, \bibfnamefont{Z}},
  \bibinfo{author}{\bibfnamefont{J.}~\bibnamefont{{Wang}}, \bibfnamefont{L}},
  \bibinfo{author}{\bibfnamefont{T.}~\bibnamefont{{Becker}}}, \bibnamefont{and}
  \bibinfo{author}{\bibfnamefont{H.}~\bibnamefont{{Walther}}},
  \bibinfo{journal}{Optics Commun.} \textbf{\bibinfo{volume}{273}},
  \bibinfo{eid}{526} (\bibinfo{year}{2007}).

\bibitem[{\citenamefont{{Rosenband} et~al.}(2008)\citenamefont{{Rosenband},
  {Hume}, {Schmidt}, {Chou}, {Brusch}, {Lorini}, {Oskay}, {Drullinger},
  {Fortier}, {Stalnaker} et~al.}}]{Rosenband08a}
\bibinfo{author}{\bibfnamefont{T.}~\bibnamefont{{Rosenband}}},
  \bibinfo{author}{\bibfnamefont{D.~B.} \bibnamefont{{Hume}}},
  \bibinfo{author}{\bibfnamefont{P.~O.} \bibnamefont{{Schmidt}}},
  \bibinfo{author}{\bibfnamefont{C.~W.} \bibnamefont{{Chou}}},
  \bibinfo{author}{\bibfnamefont{A.}~\bibnamefont{{Brusch}}},
  \bibinfo{author}{\bibfnamefont{L.}~\bibnamefont{{Lorini}}},
  \bibinfo{author}{\bibfnamefont{W.~H.} \bibnamefont{{Oskay}}},
  \bibinfo{author}{\bibfnamefont{R.~E.} \bibnamefont{{Drullinger}}},
  \bibinfo{author}{\bibfnamefont{T.~M.} \bibnamefont{{Fortier}}},
  \bibinfo{author}{\bibfnamefont{J.~E.} \bibnamefont{{Stalnaker}}},
  \bibnamefont{et~al.}, \bibinfo{journal}{Science}
  \textbf{\bibinfo{volume}{319}}, \bibinfo{pages}{1808} (\bibinfo{year}{2008}).

\bibitem[{\citenamefont{{Kleczewski} et~al.}(2012)\citenamefont{{Kleczewski},
  {Hoffman}, {Sherman}, {Magnuson}, {Blinov}, and {Fortson}}}]{Kleczewski12a}
\bibinfo{author}{\bibfnamefont{A.}~\bibnamefont{{Kleczewski}}},
  \bibinfo{author}{\bibfnamefont{R.}~\bibnamefont{{Hoffman}},
  \bibfnamefont{M}},
  \bibinfo{author}{\bibfnamefont{A.}~\bibnamefont{{Sherman}},
  \bibfnamefont{J}},
  \bibinfo{author}{\bibfnamefont{E.}~\bibnamefont{{Magnuson}}},
  \bibinfo{author}{\bibfnamefont{B.}~\bibnamefont{{Blinov}}, \bibfnamefont{B}},
  \bibnamefont{and} \bibinfo{author}{\bibfnamefont{N.}~\bibnamefont{{Fortson}},
  \bibfnamefont{E}}, \bibinfo{journal}{\pra} \textbf{\bibinfo{volume}{85}},
  \bibinfo{eid}{043418} (\bibinfo{year}{2012}).

\bibitem[{\citenamefont{{Tang} et~al.}(2013)\citenamefont{{Tang}, {Qiao},
  {Shi}, and {Mitroy}}}]{tang13a}
\bibinfo{author}{\bibfnamefont{Y.-B.} \bibnamefont{{Tang}}},
  \bibinfo{author}{\bibfnamefont{H.-X.} \bibnamefont{{Qiao}}},
  \bibinfo{author}{\bibfnamefont{T.-Y.} \bibnamefont{{Shi}}}, \bibnamefont{and}
  \bibinfo{author}{\bibfnamefont{J.}~\bibnamefont{{Mitroy}}},
  \bibinfo{journal}{\pra} \textbf{\bibinfo{volume}{87}}, \bibinfo{eid}{042517}
  (\bibinfo{year}{2013}).

\bibitem[{\citenamefont{{Gallagher}}(1967)}]{gallagher67a}
\bibinfo{author}{\bibfnamefont{A.}~\bibnamefont{{Gallagher}}},
  \bibinfo{journal}{Phys.~Rev.} \textbf{\bibinfo{volume}{157}},
  \bibinfo{pages}{24} (\bibinfo{year}{1967}).

\bibitem[{\citenamefont{{Gosselin} et~al.}(1988)\citenamefont{{Gosselin},
  {Pinnington}, and {Ansbacher}}}]{gosselin88a}
\bibinfo{author}{\bibfnamefont{R.~N.} \bibnamefont{{Gosselin}}},
  \bibinfo{author}{\bibfnamefont{E.~H.} \bibnamefont{{Pinnington}}},
  \bibnamefont{and}
  \bibinfo{author}{\bibfnamefont{W.}~\bibnamefont{{Ansbacher}}},
  \bibinfo{journal}{\pra} \textbf{\bibinfo{volume}{38}}, \bibinfo{pages}{4887}
  (\bibinfo{year}{1988}).

\bibitem[{\citenamefont{{Jin} and {Church}}(1993)}]{jin93a}
\bibinfo{author}{\bibfnamefont{J.}~\bibnamefont{{Jin}}} \bibnamefont{and}
  \bibinfo{author}{\bibfnamefont{D.~A.} \bibnamefont{{Church}}},
  \bibinfo{journal}{\prl} \textbf{\bibinfo{volume}{70}}, \bibinfo{pages}{3213}
  (\bibinfo{year}{1993}).

\bibitem[{\citenamefont{{Gerritsma} et~al.}(2008)\citenamefont{{Gerritsma},
  {Kirchmair}, {Z{\"a}hringer}, {Benhelm}, {Blatt}, and {Roos}}}]{gerritsma08a}
\bibinfo{author}{\bibfnamefont{R.}~\bibnamefont{{Gerritsma}}},
  \bibinfo{author}{\bibfnamefont{G.}~\bibnamefont{{Kirchmair}}},
  \bibinfo{author}{\bibfnamefont{F.}~\bibnamefont{{Z{\"a}hringer}}},
  \bibinfo{author}{\bibfnamefont{J.}~\bibnamefont{{Benhelm}}},
  \bibinfo{author}{\bibfnamefont{R.}~\bibnamefont{{Blatt}}}, \bibnamefont{and}
  \bibinfo{author}{\bibfnamefont{C.~F.} \bibnamefont{{Roos}}},
  \bibinfo{journal}{Eur.~Phys.~J.~D} \textbf{\bibinfo{volume}{50}},
  \bibinfo{pages}{13} (\bibinfo{year}{2008}).

\bibitem[{\citenamefont{{Safronova} and {Safronova}}(2011)}]{safronova11a}
\bibinfo{author}{\bibfnamefont{M.~S.} \bibnamefont{{Safronova}}}
  \bibnamefont{and} \bibinfo{author}{\bibfnamefont{U.~I.}
  \bibnamefont{{Safronova}}}, \bibinfo{journal}{\pra}
  \textbf{\bibinfo{volume}{83}}, \bibinfo{eid}{012503} (\bibinfo{year}{2011}).

\bibitem[{\citenamefont{{Huang} et~al.}(2011)\citenamefont{{Huang}, {Liu},
  {Cao}, {Ou}, {Liu}, {Guan}, {Huang}, and {Gao}}}]{huang11a}
\bibinfo{author}{\bibfnamefont{Y.}~\bibnamefont{{Huang}}},
  \bibinfo{author}{\bibfnamefont{Q.}~\bibnamefont{{Liu}}},
  \bibinfo{author}{\bibfnamefont{J.}~\bibnamefont{{Cao}}},
  \bibinfo{author}{\bibfnamefont{B.}~\bibnamefont{{Ou}}},
  \bibinfo{author}{\bibfnamefont{P.}~\bibnamefont{{Liu}}},
  \bibinfo{author}{\bibfnamefont{H.}~\bibnamefont{{Guan}}},
  \bibinfo{author}{\bibfnamefont{X.}~\bibnamefont{{Huang}}}, \bibnamefont{and}
  \bibinfo{author}{\bibfnamefont{K.}~\bibnamefont{{Gao}}},
  \bibinfo{journal}{\pra} \textbf{\bibinfo{volume}{84}}, \bibinfo{eid}{053841}
  (\bibinfo{year}{2011}).

\bibitem[{\citenamefont{{Huang} et~al.}(2012)\citenamefont{{Huang}, {Cao},
  {Liu}, {Liang}, {Ou}, {Guan}, {Huang}, {Li}, and {Gao}}}]{huang12a}
\bibinfo{author}{\bibfnamefont{Y.}~\bibnamefont{{Huang}}},
  \bibinfo{author}{\bibfnamefont{J.}~\bibnamefont{{Cao}}},
  \bibinfo{author}{\bibfnamefont{P.}~\bibnamefont{{Liu}}},
  \bibinfo{author}{\bibfnamefont{K.}~\bibnamefont{{Liang}}},
  \bibinfo{author}{\bibfnamefont{B.}~\bibnamefont{{Ou}}},
  \bibinfo{author}{\bibfnamefont{H.}~\bibnamefont{{Guan}}},
  \bibinfo{author}{\bibfnamefont{X.}~\bibnamefont{{Huang}}},
  \bibinfo{author}{\bibfnamefont{T.}~\bibnamefont{{Li}}}, \bibnamefont{and}
  \bibinfo{author}{\bibfnamefont{K.}~\bibnamefont{{Gao}}},
  \bibinfo{journal}{\pra} \textbf{\bibinfo{volume}{85}}, \bibinfo{eid}{030503}
  (\bibinfo{year}{2012}).

\bibitem[{\citenamefont{{Berkeland} et~al.}(1998)\citenamefont{{Berkeland},
  {Miller}, {Bergquist}, {Itano}, and {Wineland}}}]{berkeland98a}
\bibinfo{author}{\bibfnamefont{D.~J.} \bibnamefont{{Berkeland}}},
  \bibinfo{author}{\bibfnamefont{J.~D.} \bibnamefont{{Miller}}},
  \bibinfo{author}{\bibfnamefont{J.~C.} \bibnamefont{{Bergquist}}},
  \bibinfo{author}{\bibfnamefont{W.~M.} \bibnamefont{{Itano}}},
  \bibnamefont{and} \bibinfo{author}{\bibfnamefont{D.~J.}
  \bibnamefont{{Wineland}}}, \bibinfo{journal}{J.~Appl.~Phys.}
  \textbf{\bibinfo{volume}{83}}, \bibinfo{pages}{5025} (\bibinfo{year}{1998}).

\bibitem[{\citenamefont{{Deutsch} and {Jessen}}(2010)}]{deutsch10a}
\bibinfo{author}{\bibfnamefont{I.~H.} \bibnamefont{{Deutsch}}}
  \bibnamefont{and} \bibinfo{author}{\bibfnamefont{P.~S.}
  \bibnamefont{{Jessen}}}, \bibinfo{journal}{Optics~Commun.}
  \textbf{\bibinfo{volume}{283}}, \bibinfo{pages}{681} (\bibinfo{year}{2010}).

\bibitem[{\citenamefont{{Arora} et~al.}(2007)\citenamefont{{Arora},
  {Safronova}, and {Clark}}}]{arora07a}
\bibinfo{author}{\bibfnamefont{B.}~\bibnamefont{{Arora}}},
  \bibinfo{author}{\bibfnamefont{M.~S.} \bibnamefont{{Safronova}}},
  \bibnamefont{and} \bibinfo{author}{\bibfnamefont{C.~W.}
  \bibnamefont{{Clark}}}, \bibinfo{journal}{\pra}
  \textbf{\bibinfo{volume}{76}}, \bibinfo{eid}{052509} (\bibinfo{year}{2007}).

\end{thebibliography}

\end{document}